\documentclass[prl,10pt, superscriptaddress, twocolumn]{revtex4}
\usepackage[intlimits]{amsmath}
\usepackage{amssymb}
\usepackage{graphicx}
\usepackage{graphics}
\usepackage{epstopdf}
\usepackage{dcolumn}
\usepackage{hyperref}
\usepackage{natbib}
\begin{document}
\title{Rippling Instabilities in Suspended Nanoribbons}

\author{Hailong Wang}
\affiliation{Group for Simulation and Theory of Atomic-Scale Material Phenomena ({\it st}AMP), Department of Mechanical and Industrial Engineering, Northeastern University, Boston MA 02115}
\author{Moneesh Upmanyu}
\affiliation{Group for Simulation and Theory of Atomic-Scale Material Phenomena ({\it st}AMP), Department of Mechanical and Industrial Engineering, Northeastern University, Boston MA 02115}

\begin{abstract}
Morphology mediates the interplay between the structure and electronic transport in atomically thin nanoribbons such as graphene as the relaxation of edge stresses occurs preferentially via out-of-plane deflections.
%similar in principle to the wavy edges that decorate flowers, leaves and ripped plastic sheets. 
In the case of end-supported suspended nanoribbons that we study here, past experiments and computations have identified a range of equilibrium morphologies, in particular for graphene flakes, yet a unified understanding of their relative stability remains elusive. Here, we employ atomic-scale simulations and a composite framework based on isotropic elastic plate theory to chart out the morphological stability space of suspended nanoribbons with respect to intrinsic (ribbon elasticity) and engineered (ribbon geometry) parameters, and the combination of edge and body actuation. The computations highlight a rich morphological shape space that can be naturally classified into two competing shapes, bending-like and twist-like, depending on the distribution of ripples across the interacting edges. The linearized elastic framework yields exact solutions for these rippled shapes. For compressive edge stresses, the body strain emerges as a key variable that controls their relative stability and in extreme cases stabilizes co-existing transverse ripples. Tensile edge stresses lead to dimples within the ribbon core that decay into the edges, a feature of obvious significance for stretchable nanoelectronics. The interplay between geometry and mechanics that we report should serve as a key input for quantifying the transport along these ribbons.
\end{abstract}
%\pacs{62.25.+g, 85.85+j, 85.35.Kt, 81.07.De, 62.20.Dc}
%\keywords{carbon nanotube networks; fluidic assembly; topology; electronic transport}
\maketitle

\section{Introduction}
The performance of nanoelectronic devices based on atomically thin films such as graphene depends critically on the interplay between geometry, structure and mechanics. This is especially true in the case of nanoribbons where the edge structure can fundamentally alter the overall response. As a classic example, the band gap in graphene nanoribbons (GNRs) is sensitive to edge type and ribbon width and in extreme cases determines the nature of the electronic transport, metallic or semiconducting~\cite{nr:TerronesTerrones:2010, nr:Nakada:1996, nr:Ezawa:2006, gnr:ChenAvoruris:2007, gnr:HanKim:2007, nr:GunlyckeWhite:2010}. The interplay with ribbon morphology serves as a crucial ingredient in quantifying these structure-property relations as the edge elasticity varies significantly with its structure~\cite{nr:ReddyShenoyZhang:2009, gph:GanSrolovitz:2010}. The edge shape also determines the nature and extent of edge functionalization~\cite{nr:DuboisRoche:2010, gph:UthaisarBarone:2009}. Past studies on these ultra-thin ribbons and sheets have revealed several interesting morphologies. In instances where the edge stress is compressive, the edges warp out of plane; Shenoy and coworkers used a combination of atomic-scale computations and scaling arguments to calculate the wavelength, amplitude and penetration width of such undulations in semi-infinite edges~\cite{nr:ShenoyZhang:2008}. Independent computations by Bets and Yakobson show that the rippling wavelength scales with the ratio of the edge stress to the flexural rigidity $\tau_e/D$~\cite{nr:BetsYakobson:2010}. Below a critical width, the ripples transition into a spontaneous twist. The situation is expectedly different for tensile edge stresses, recently observed in reconstructed edges in graphene and intrinsic to bilayer (and possibly multi-layer) ribbons that reconstruct into partial edge tubules~\cite{nr:ShenoyZhang:2010, nr:HuangYakobsonLi:2009}. Here, the out-of-plane displacement occurs preferentially away from the edge such that the ribbon midsection curls as it ripples~\cite{nr:ShenoyZhang:2010}.

Our focus here is on end-supported suspended nanoribbons shown schematically in Fig.~\ref{fig:fig1}a. Unlike freestanding nanoribbons, this architectural motif is a natural building block for next-generation nanoelectronic devices and NEMS devices as it allows controlled yet scalable device integration while minimizing deleterious substrate effects~\cite{nr:TerronesTerrones:2010, gnr:ChenAvoruris:2007, gnr:HanKim:2007, gnr:BolotinKimStormer:2008, gnr:WangDai:2008, gnr:ChenKimHone:2009}. The ribbons can be trimmed to shape before or after clamping them onto the end-supports (electrodes)~\cite{gnr:LiDai:2008, gnr:DattaJohnson:2008, gnr:CiAjayan:2008, gnr:LambinBiro:2008, gnr:BaiHuang:2009, gnr:KosynkinTour:2009, gnr:JiaoDai:2009, gnr:WuCheng:2010}, and the two scenarios result in differing mechanical constraints on the ribbons, as detailed later in this article. Little is known regarding their morphological stability for several reasons. One, the edge stress-induced morphologies reported in past studies have been analyzed primarily to understand post-buckled shapes, via atomic-scale computations or scaling analyses with attendant simplifications~\cite{nr:ShenoyZhang:2008,nr:ShenoyZhang:2010}. However, a detailed understanding of the pre-buckled shapes is necessary. These become tractable only in the (linear) small amplitude limit and a naive approach would discount their utility as the post-buckling in these atomically thin sheets is expected to take place primarily through bending - the stretching is prohibitively expensive~\cite{memb:SharonSwinney:2002, memb:MarderSmith:2003, memb:AudolyBoudaoud:2003, memb:SchrollDavidovitch:2011}. This is in stark contrast to recent computations on edge morphologies of semi-infinite graphene sheets that show that stretching plays crucial, if not decisive role~\cite{nr:ShenoyZhang:2008}. Furthermore, as we show below, the computed buckled shapes are quite sensitive to the initial perturbations, even more so for confined systems such as end-supported nanoribbons. Then, the stability of the precursory small amplitude deformations, where both stretching and bending can influence the stability, is the key to understanding the post-buckled shapes. Two, the end-supports as well as the edge-edge interactions limit the possible morphologies as they force to ribbon to be on average flat. More specifically, global buckling modes such as (developable) twist or saddle shapes become untenable and edge stress accommodation takes place primarily through periodic ripples that also interact across the edges. Lastly, reconciliation of the shapes observed in experiments (for e.g.~\ref{fig:fig1}b and~\ref{fig:fig1}c) or atomic-scale simulations with elastic theories must  factor in thermal effects that naturally manifest as shape fluctuations. Statistical theories of polymerized elastic membranes can be invoked to account for these effects via renormalized ribbon elasticity~\cite{memb:NelsonPiranWeinberg:1992, gph:DoussalRadzihovsky:1992}. While we do not address this specific issue here, it must be noted that the highly strain-sensitive bonding that stabilizes these flakes also results in strongly size-dependent and non-classical effects 
%(for e.g., intrinsic rippling reported in graphene) 
which can become significant at these scales~\cite{gph:FasoliniKatsnelson:2007}. 

In this article, we use stability analyses and computations to explore the morphological stability space of nanoribbons as a function of both intrinsic and engineered parameters, i.e. ribbon geometry (width $w$ and length $l$), material parameters (sheet and edge stiffness $S$ and $S_e$, and $\tau_e$) and deformation along the ribbon (x-)axis (uniaxial strain $\epsilon_{xx}$). Atomic-scale simulations of ribbons serve as inputs for identifying the possible equilibrium shapes. Although the computations focus exclusively on GNRs, we demonstrate the generality of these results via rigorous stability analyses based on isotropic elastic plate theory, which in turn allow us to develop stability diagrams for a combination of geometric, material and processing parameters. %In each case, systematic atomic-scale computations are then employed to validate the theoretical results. 
%We conclude with a discussion on thermal effects via simple scaling arguments. 

\section{Past Experiments and Computations}
\begin{figure*}[htb]
\centering
\includegraphics[width=1.5\columnwidth]{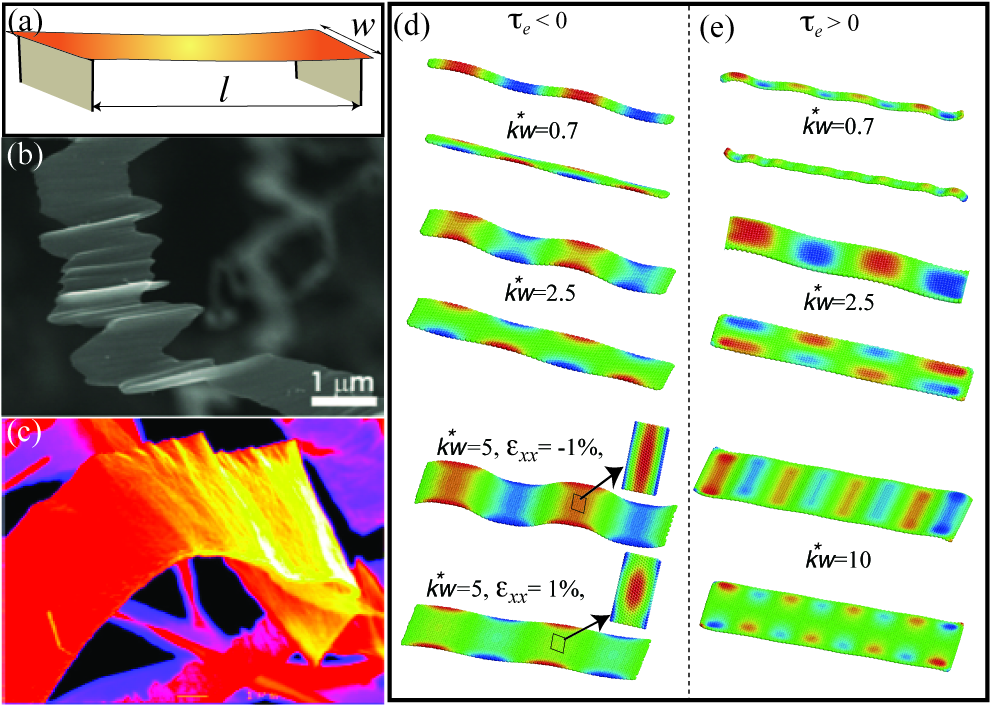}
\caption{\label{Shape} (a) Schematic illustration of a suspended nanoribbon, simply supported or clamped depending on the end-supports. Experimental observations of (b) rippling in multilayer graphene ribbons~\cite{nr:Campos-DelgadoTerrones:2008} and (c) edge rippling in hollow BN nanoribbons~\cite{nr:ChenChengLu:2008}. The images are reproduced with the authors' permissions. (d, e) Similar shapes observed in atomic-scale computations on $l\approx20-25$\,nm long, armchair-terminated graphene nanoribbons (a-GNRs) with (d) compressive and (e) tensile edge stresses. The critical wavenumber $k^\ast w$ is the ratio of the ribbon width to ripple wavelength. In all cases, both bending-like (symmetric, S) and twist-like (antisymmetric, AS) shapes are observed. The color indicates the scaled magnitude of the out-of-plane displacements as an indicator of the extent of edge-edge interactions. The additional effect of applied compressive and tensile strains  for $\tau_e<0$ is shown in bottom two plots in (d). The insets show details of the bent morphology at and around the ribbon mid-line.}
\label{fig:fig1}
\end{figure*}

\ref{fig:fig1} showcases the periodic ripples that have been observed in suspended nanoribbons. The experimental images (\ref{fig:fig1}b and~\ref{fig:fig1}c) show two differing morphologies, bending-like rippling in multi-layered graphene ribbons~\cite{nr:Campos-DelgadoTerrones:2008} and edge rippling in hollow BN ribbons~\cite{nr:ChenChengLu:2008}. The results of our atomic-scale simulations on graphene nanoribbons, summarized in~\ref{fig:fig1}c, reveal a considerably richer morphological space. To facilitate direct comparison with past computational studies, we have performed these computations using a reactive bond order (AIREBO) potential~\cite{pot:Stuart:2000} as implemented in the software package LAMMPS~\cite{pot:Plimpton:1995} (see Methods). For each combination of parameters, the characteristics associated with the rippling are extracted from the relaxed shapes in the computations. The ratio of the width to wavelength $\lambda$ associated with the rippled shape is indicated in the figure, expressed as a dimensionless wavenumber $kw$, where $k=2\pi/\lambda$. While we do not discuss the effect of ribbon length $l$ in detail in the remainder of the article, its main effect is to act as a constraint on the permissible wavenumbers in instances where $\pi/l>k$.

Some of the shapes we observe are similar to those reported in past studies, both at zero and finite temperature~\cite{nr:ShenoyZhang:2008, nr:BetsYakobson:2010, nr:ShenoyZhang:2010, nr:LuHuang2010}. A key feature is that we observe two dominant morphological classes for each given set of parameters: in-phase or symmetrical ripples (S), and out-of-phase or asymmetrical (AS) ripples, where the phase refers to the relative displacements of the edges. The specific form of ripples is controlled by the sign of the edge stress and the aspect ratio. For compressive stresses and sufficiently large widths, the edges do not interact and the ribbon exhibits classical edge ripples (not shown). As the width is reduced, the edge-edge interactions become important such that the ripples penetrate through the width and ribbon buckles either in- or out-of-phase~\cite{nr:LuHuang2010}. At much smaller widths, the entire ribbon buckles with bending-like or twist-like undulations. Applied longitudinal strains $\epsilon_{xx}$ modify the midline-line morphology, as shown in the figure for symmetrically rippled ribbons. Tensile edge stresses force the midline to curl out-of-plane. However, since end-conditions require the midline to be flat, we observe dimples at and around the midline which decay into the edges. The dimples can also split asymmetrically at large widths, although the mode is relatively rare. At small widths, though, the ribbon morphology is either flat or symmetrically rippled. 
%For a fixed set of parameters, the initial random perturbations also influence the form of the rippled morphology, as expected.

We note that some of these shapes, in particular symmetric ripples in ribbons with compressed edges (e.g. filament-like buckling at small widths) are also observed in naturally occurring ribbons such as straight-edged long leaves~\cite{memb:MarderSmith:2003, elastica:Marder:2003, elastica:LiangMahadevan:2009}. While there are parallels with the rippled shapes analyzed here, the shapes in the materially homogeneous natural systems are driven by inelastic and distributed growth strains unlike the highly localized and elastically stressed edges that characterize the nanoribbons considered here. Coupled with the differing end-conditions that arise in suspended nanoribbons, the stability of the expected shapes is qualitatively different.

%We first perturb the atoms of graphene nanoribbon by allowing symmetric out-of-plane displacement $\zeta=A\sin{kx}$ and antisymmetric out-of-plane displacement $\zeta=2Ax\sin{kx}/w$, the amplitude $A=0.0nmA$. Then the atoms in the perturbed nanoribbon are then allowed to relax using a conjugate gradient algorithm implemented in LAMMPS\cite{Plimpton1995} with an energy tolerance of $10^{-10}$. Graphene ribbons with edge buckling are shown in Fig. \ref{Shape}, the color indicates the scaled magnitude of out-of-plane displacements of the atoms $\zeta/|f(w/2)|$.

\section{Stability Analysis of Periodic Ripples}
%\subsection{Governing Equations}
The diverse morphologies observed in the computations are systematically analyzed using classical elastic plate theory. Consider a nanoribbon of thickness $h$ and width $w$ ($h\ll w$) clamped to supports spaced apart by a length $l$. We assume that the ribbon is a linear, isotropic elastic thin plate with elastic modulus $E$, Possion's ratio $\nu$, and bending and stretching stiffnesses $D=Eh^3/12(1-\nu^2)$ and $S=Eh$, respectively. The edge stress $\tau_e$ in these atomically thin nanoribbons arises due to structural changes or reconstructions localized at the edges. In continuum limit, then, the edge can be approximated as an elastically stressed bounding spring with negligible bending stiffness~\cite{book:TimoshenkoGere:1961}. This composite approximation is similar in principle to the core-shell framework often invoked to describe elastic behavior or nanowires and thin films~\cite{nw:MillerShenoy:2000, nw:HaiyiUpmanyuHuang:2005, nw:WangUpmanyu:2008}. For a system so structured strain compatibility and force equilibrium require that the elastic Hamiltonian that maps the initially flat ribbon to its deformed state, ${\bf R}\equiv(x, y, 0)$ $\rightarrow$ ${\bf R^\prime}\equiv(x+u_x, y + u_y, \zeta)$, satisfies the generalized F\"{o}ppl-von K\'{a}rm\'{a}n (F-vK) equations~\cite{book:TimoshenkoGere:1961}. 
The stability of the periodic ripples can be analyzed by assuming a sinusoidal variation in the out-of-plane deflection, 
\begin{equation}
\label{eq:sinusoidalPert}
\zeta(x,y)=f(y)\sin{kx}.
\end{equation}
%where the ripple wave number is the inverse wavelength and is also related to its length $l$, $k=2\pi/\lambda=n\pi/l$. 
The shape satisfies the boundary conditions along the simply supported sides of the graphene nanoribbon since $\zeta=0$ and $\zeta_{,xx} +\nu\zeta_{,yy}=0$ for  $x=0$ and $x=l$. In the limit of negligible transverse and shear stresses, the classical F-vK equations simplify to a boundary value problem for the out-of-plane deflection (see Methods),
\begin{eqnarray}
\label{von1_ripple}
f_{,yyyy}-2k^2f_{,yy}+\left( k^4+k^2 \epsilon_{xx} \frac{S}{D} \right) f &=&0,
\end{eqnarray}
where $\epsilon_{xx}=\epsilon_{xx}^0 + \epsilon_{xx}^{a}$ is a general uniaxial strain, expressed as the sum of an intrinsic strain $\epsilon_{xx}^0$, if present, and applied uniaxial strain $\epsilon_{xx}^a$. Together with the boundary conditions,
\begin{eqnarray}
\label{eq:rippledBC}
(f_{,yy} - \nu k^2 f)|_{\pm{\frac{w}{2}}}=0\nonumber\\
\{[f_{,yyy}-(2-\nu) k^2f_{,y}]\mp\frac{\tau_e}{D}k^2f \} |_{\pm{\frac{w}{2}}}=0,
\end{eqnarray}
we completely specify the form of the deflection. 
\begin{figure*}[htb]
\centering
\includegraphics[width=1.6\columnwidth]{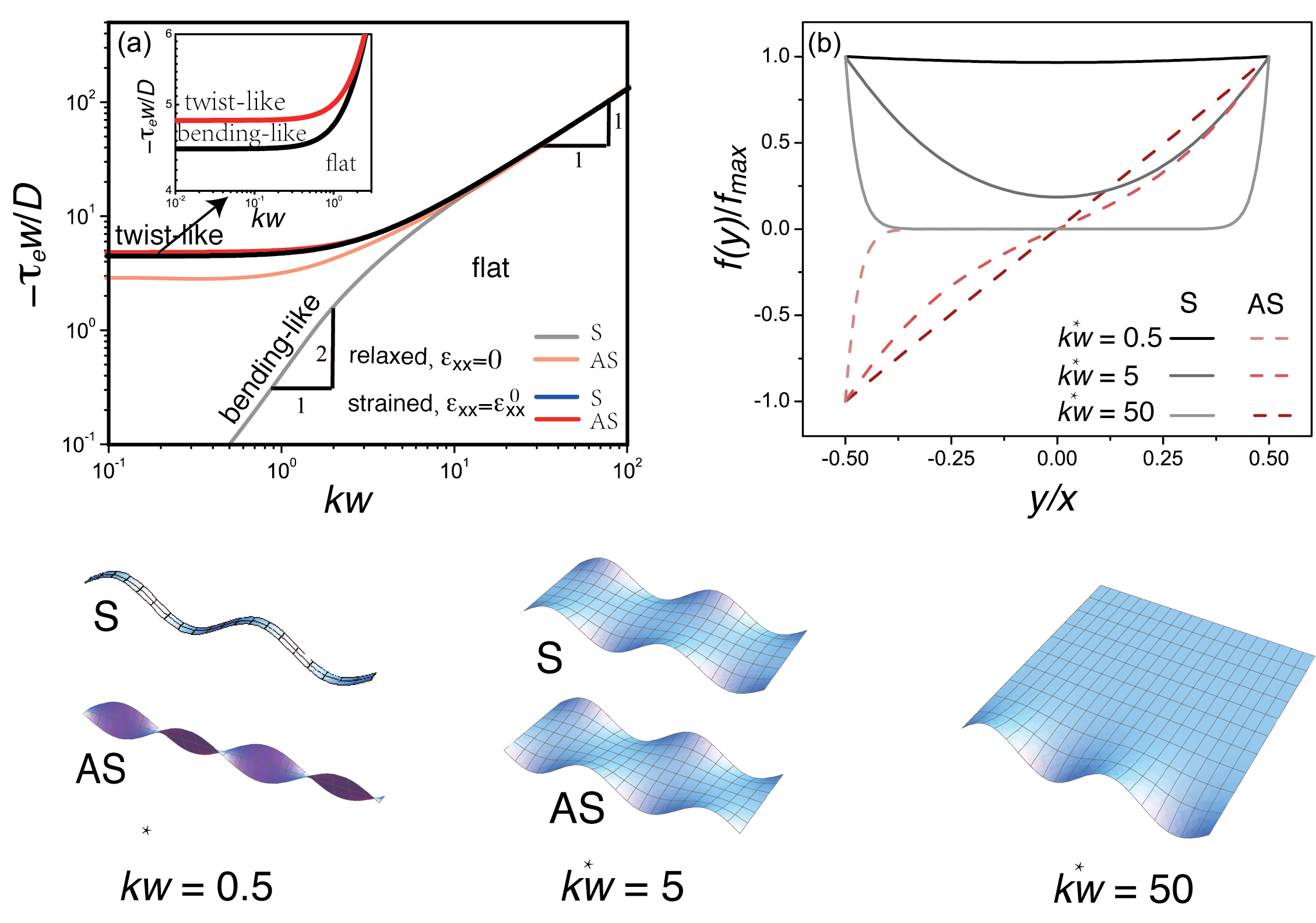}
\caption{\label{fig:fig2_1} (a) Morphological stability diagram showing the scaled edge stress ${\tau_e}w/D$ as a function of the wave number $kw$ for relaxed ($\epsilon_{xx}=0$, light shading) and intrinsically strained ($\epsilon_{xx}^0=-2\tau_e/(Sw)$, dark shading)) nanoribbons with compressive edge stresses. Here as well as in the following figures the critical curves for bending-like and twist-like buckling are indicated by black and red lines, respectively. (inset) Magnified plot for small wavenumbers $kw$ that shows the bifurcation between the two morphological classes for strained ribbons.  (b) The ribbon profile obtained from the analytical solution (see SI) for relaxed ribbons plotted as the scaled deflection $f(y)/f_{max}$ versus scaled width $y/w$ for the three different wavenumbers. For both classes, the rippling localizes to the edges with increasing wavenumber. Schematic illustrations of the shapes for some of the profiles are also shown. All plots are based on Poisson's ratio $\nu=0.17$ corresponding to that for graphene~\cite{nr:Blakslee:1970}.
}
\end{figure*} 

The effect of the residual edge stress $\tau_e=\tau_e^0 + S_e\epsilon_{xx}$ that enters into the boundary value problem depends on the synthesis procedure. Here, $\tau_e^0$ is the unrelaxed edge stress. In cases where the ribbons are very long, or trimmed to shape following synthesis and then transferred onto the end supports, the edge stresses result in a residual longitudinal force that must be borne by the ribbon core. A simple force balance for the initially flat ribbon yields the intrinsic body strain, $2\tau_e + Sw\epsilon_{xx}^0=T\approx0$, where $T$ is the net longitudinal force. The body strain acts much like an imposed uniaxial strain over the entire ribbon core and therefore is analyzed in the context of extrinsically strained ribbons. An entirely different scenario occurs when the precursor flake is placed on the end-supports and then trimmed to shape. In this case, the edge-supports modify the net force balance such that the ribbon core is unstrained, $\epsilon_{xx}=0$.

%As expected, the ratio of the edge stress to the bending stiffness $\tau_e/D$ emerges as a key component of the end conditions for the boundary value problem.

\begin{figure*}[htb]
\centering
\includegraphics[width=1.6\columnwidth]{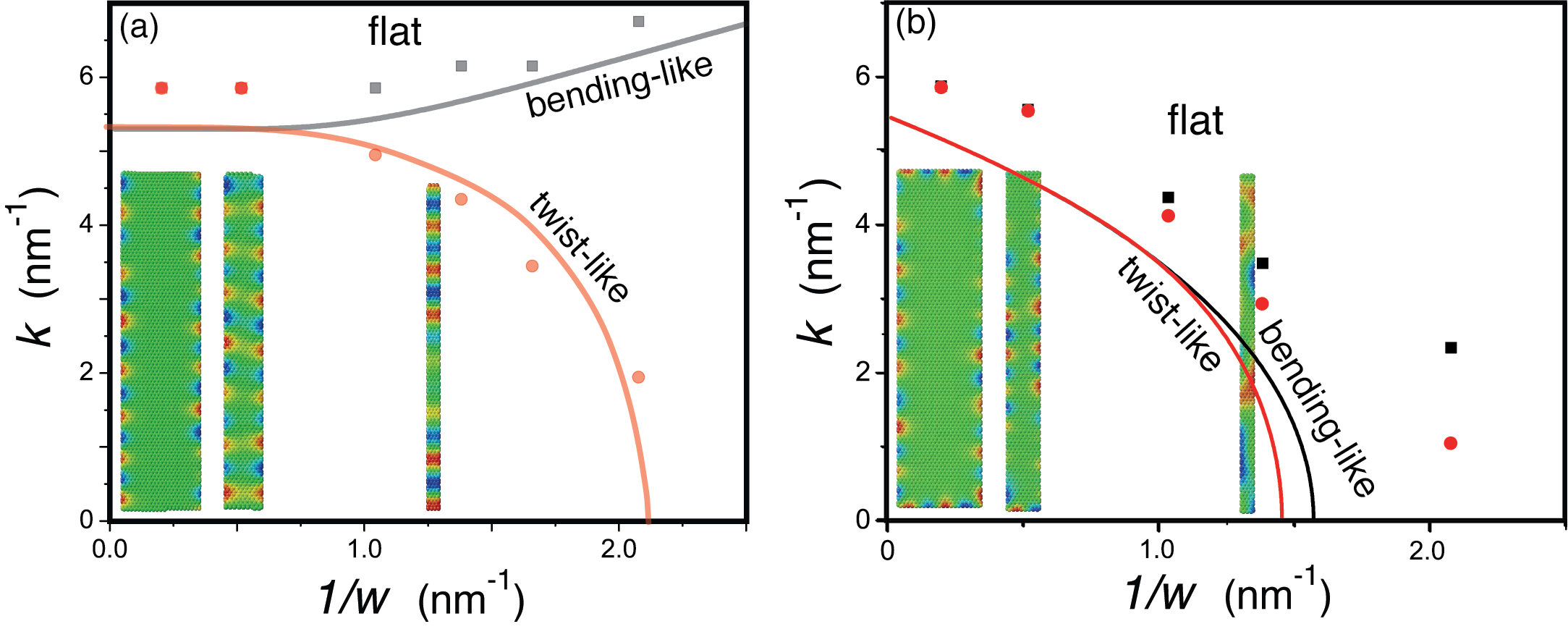}
\caption{\label{fig:fig2_2} (a-b) Comparison of the analytically predicted stability of of bending-like and twist-like shapes and the buckled shapes observed in atomic-scale simulations for (c) relaxed and (d) intrinsically strained a-GNRs, now plotted as $k(w)$. The representative atomic-configurations for symmetrically rippled ribbons with widths $w=0.72$\,nm, $w=1.93$\,nm and $w=5.12$\,nm are also shown (inset).
}
\end{figure*}

\section{Compressive Edge Stresses}
\subsection{Relaxed Ribbons}
We first analyze relaxed ribbons characterized by $\epsilon_{xx}=0$ and $\tau_e<0$. The boundary conditions (~\ref{eq:rippledBC}) yield the critical point for the onset of buckling, conveniently expressed as stability diagrams that relate the (dimensionless) edge stress $\tau_e w/D$ to the scaled wavenumber $kw$.  \ref{fig:fig2_1}a shows these stability diagrams for both bending-like (S) and twist-like (AS) ripples in a ribbon with Poisson's ratio corresponding to that for graphene, $\nu=0.17$. The critical stress varies linearly both morphological classes at large widths (or equivalently short ribbons), $\tau_e^\ast w/D\approx-[(1-\nu)(3+\nu)] (k^\ast w)/2$ for $k^\ast w\gg 1$. The solution also yields the ribbon shape (SI, Eqs. S4 and S6), plotted in~\ref{fig:fig2_1}b as a shape function $f(y)/f_{max}$ for several representative wavenumbers. For small aspect ratios (e.g. $k^\ast w=50$), the ripples are uncorrelated and localized to the edges, as expected. At the critical point $k^\ast w\approx 5$, we see a bifurcation due to edge ripples that now begin to interact across the width via saddle-shapes morphologies with net negative Gaussian curvature, apparent in the schematic illustrations in~\ref{fig:fig2_1}b and also in the post-buckled shapes observed in computations (\ref{fig:fig1}d, with $kw=2.5$).

The asymptotic behavior for the limit $kw\ll1$ sheds light on the markedly different behavior for the two morphological classes. The bending-like ripples exhibit a quadratic dependence, $\tau_e^\ast w/D\approx-(1-\nu)(1+\nu)(k^\ast w)^2$ while the critical edge stress for twist-like ripples is independent of the wavenumber, $\tau_e^\ast w/D\approx-4(1-\nu)$. As an example, the almost flat (scaled) profile of a high aspect ratio ribbon with $k^\ast w=0.5$ is plotted in~\ref{fig:fig2_1}b. The stability curve for the shape also follows from simple scalings based on elastic energies (calculated per ripple wavelength $\lambda$) associated with ribbon bending $\mathcal{E}_b$, and ribbon stretching at the core and edge, $\mathcal{E}_s$ and $\mathcal{E}_s^e$. The bending energy follows from the curvature tensor, $ \mathcal{E}_b \sim Dkw\int^\lambda_0 (\kappa_{xx}^2+\kappa_{yy}^2) \,dx $ and $ \mathcal{E}_b \sim Dkw\int^\lambda_0 \kappa_{xy}^2 \,dx $ for the bending-like and twist-like ripples, respectively.
%$\mathcal{E}_b=D(\kappa_{xx}^2+2\kappa_{xy}^2+\kappa_{yy}^2) w$. 
The stretching energies are related to longitudinal and edge strains, $\mathcal{E}_s \sim -Tkw\int^\lambda_0  \epsilon_{xx} \,dx$ and $\mathcal{E}_s \sim -2\tau_e kw\int^\lambda_0  \epsilon_{xx} \,dx$.
%$\mathcal{E}_s =\int -(S\epsilon_{xx}) \epsilon_{xx}^0 w\,dx$ and $\mathcal{E}_s^e\sim -\tau\epsilon_e$. 
Ignoring the weak variation in out-of-plane deflection across the width, $f\sim\delta_x$ and $f\sim2y/w\delta_x$ for the bending-like and twist-like ripples, respectively. Then, $\kappa_{xx} \sim -k^2 \delta_x^2 \sin{kx}$ and $\kappa_{yy} \sim 0$ and the dominant contributions for bending-like ripples scale as $\mathcal{E}_b\sim Dk^4\,\delta_x^2w$ and $\mathcal{E}_s^e\sim-\tau_ek^2\delta_x^2$. Taken together, they yield the quadratic dependence, $-\tau_e^\ast w/D{\sim} (k^*w)^2$. In the case of twist-like ripples, the stretching energy remains unchanged. The bending energy is due to $\kappa_{xy} \sim k\delta_x/w \cos{kx}$ and is relatively larger, $\mathcal{E}_b\sim Dk^2/w$. Equating the two yields the critical edge stress, $-\tau_e^\ast w/D{\sim}1$.

The stability diagram yields insight into the shapes observed in atomic-scale simulations on GNRs. For unreconstructed armchair and zigzag terminations, the compressive edge stresses range from $\tau_e^0=-10.5$\,eV/nm to $-20.5$\,eV/nm while the edge stiffness $S_e$ varies from $113$ to $147$\,eV/nm~\cite{nr:ReddyShenoyZhang:2009, gph:GanSrolovitz:2010}. The low bending stiffness of these atomically thin sheets ($D\approx 1.5$\,eV) yields a scaled edge stress $\tau_ew/D\sim10$ for nanometer-wide GNRs. The critical wavenumber is in the vicinity of the bifurcation point where both bending- and twist-like undulations are possible, although the former are energetically favored. This is corroborated by the computed morphologies for a-GNRs with varying widths, shown in~\ref{fig:fig1}d-e  and~\ref{fig:fig2_2}a (inset). Both morphological classes are observed depending on the form of the  perturbation. %highlighting the dynamical nature of the stability and its sensitivity to the initial conditions.
\ref{fig:fig2_2}a shows the reduced stability diagram, $k$ vs. $w$ predicted by our analysis and that extracted from the computations. The overall trend is well-described by our stability diagram although there are quantitative deviations in the simulated shapes. They are likely due to the continuum approximation of an atomic-scale system and the fact that we ignore the relaxation of the edge stress due to out-of-plane displacements~\cite{gph:GanSrolovitz:2010}. Furthermore, a more detailed analysis of the atomic-configurations reveals that at intermediate to large widths, the two morphological classes can also co-exist, albeit with differing ripple wavelengths. This can be clearly seen in the atomic-configuration for the $w=1.93$\,nm a-GNR shown in Fig.~\ref{fig:fig2_2}a (inset). We see an out-phase component in the symmetric ripples away from the ribbon ends, indicative of a twist-like buckled mode with a much longer wavelength that allows the ribbon to further relax the residual edge stress. At very small widths, the co-existence shapes lead to locally flat morphology quite like the one shown for $w=0.72$\,nm. 
%Finally, the simulations clearly indicate that one cannot ignore initial conditions on the shape stability. For example, 
%since the dynamic stability is expected to be sensitive to the initial conditions during the evolution into post-buckled shapes. . 

\subsection{Strained Ribbons} The eigenvalue solution can be written as
\begin{equation}
\Phi(kw,{\tau_e}w/D,\epsilon_{xx}Sw^2/D)=0,
\end{equation}
where $\epsilon_{xx}Sw^2/D$ is the additional (scaled) strain. As a starting point, we explore the effect of intrinsic body strains that arise naturally in long nanoribbons, $\epsilon_{xx}=\epsilon_{xx}^0$. The compressive edge stresses require that $\epsilon_{xx}=-2\tau_e/S>0$. The modified stability stability diagram, also shown in~\ref{fig:fig2_1}a, is based on a lengthy analytical solution which is not shown for brevity (see SI).
%\begin{align}
%\label{result_positive}
%f&=A_0\cosh{py}\cos{qy}+A_1\cosh{py}\sin{qy}\nonumber\\
%&+B_0\sinh{py}\sin{qy}+B_1\sinh{py}\cos{qy},
%\end{align}
%where the ratio $\xi=k_x/k$ defines the variables $p$ and $q$, i.e. $2p^2=k^2(\sqrt{1+\xi^2} + 1)$ and $2q^2=k^2(\sqrt{1+\xi^2}-1)$. As a check, we recover the relaxed solution ($\epsilon_{xx}=0$, Eq.~\ref{result_zero}) since $(\cosh{py}, \sinh{py}, \cos{qy}, \sin{qy})$ $\rightarrow$ $(\cosh{ky}, \sinh{ky}, 1, ky)$  in the limit $k_x{\ll}k$. 
%shows the modified stability diagram for a ribbon with a stretching stiffness corresponding to graphene, $S=2000$\,eV/nm$^2$. 
It is immediately clear that the flat phase is more stable as the edge-edge interactions can be absorbed for relatively wider ribbons due to the mediating strained core. In effect, the relaxation of the edge ripples shifts to smaller wavenumbers and larger edge stresses as it now occurs with respect to an already stretched ribbon core. The bifurcation into the two morphological classes shifts accordingly (inset). Of importance is the dramatic effect of the intrinsic strain on the relative stability of the bending- and twist-like ripples for small aspect ratio ribbons, $kw\ll1$. Although the critical point for bending-like ripples is still lower (inset), we no longer see the quadratic dependence in the critical point for bending-like ripples. Rather, quite like the twist-like rippling in relaxed ribbons, both morphological classes result in rippled states that are determined entirely by the (larger) critical stresses, $\tau_e^\ast w/D$, that are in turn independent of the wavenumber. The fact that twist-like ripples compete effectively is not surprising as they preserve the midline length and therefore are clearly efficient in accommodating the tensile strain within the core. In the specific case of a-GNRs, the predicted stability space $k(w)$ as well as the results of the computations are plotted in~\ref{fig:fig2_2}b. As in the unstrained case, the analytical framework captures the trends although the quantitative agreement becomes noticeably poor at small widths due to effects mentioned earlier.
%and also due to inaccuracies associated with the core-spring framework that we employ here. 
\begin{figure*}[thb]
\centering
\includegraphics[width=1.4\columnwidth]{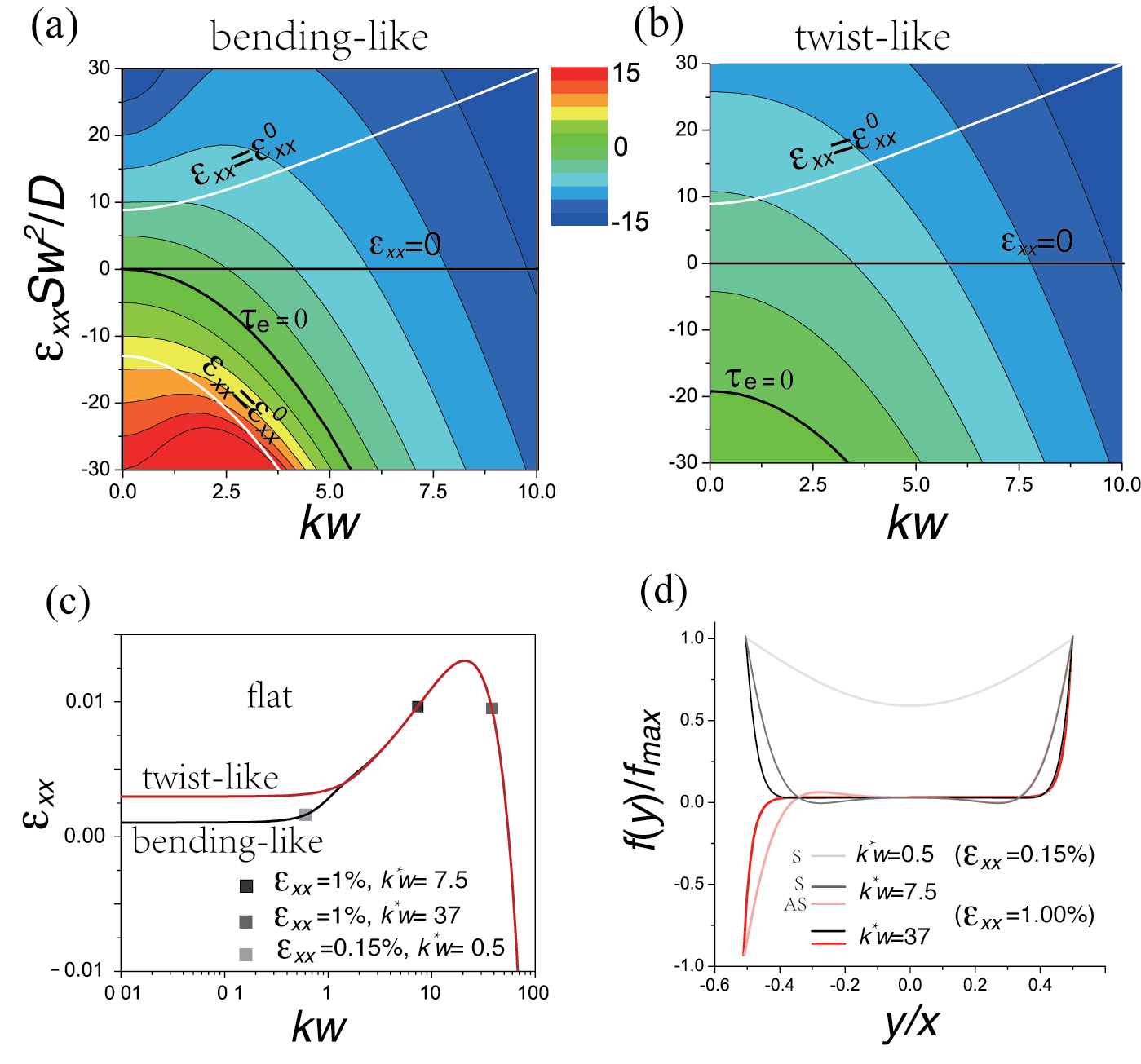}
\caption{(a-b) Iso-(edge)stress critical contours for (a) bending-like and (b) twist-like rippling of strained ribbons with compressive and tensile edge stresses. The contour plot associated with classical Euler buckling ($\tau_e=0$) is also plotted. The solid white lines are iso-strain plots for intrinsically strained nanoribbons summarized in~\ref{fig:fig2_1} and \ref{fig:tensileEdgeStress}a. (c) The critical curves for strained unreconstructed a-GNRs with compressive edge stress $\tau_e=10.5$\,eV/nm. (d) Ribbon profiles $f(y)/f_{max}$ for some of the critical points below ($\epsilon_{xx}=0.15\%$, $k^\ast w=0.5$) as well as in the vicinity of the peak strain ($\epsilon_{xx}=1\%$, $k^\ast w=7.5, 37$).
\label{fig:fig3}
}
\end{figure*}

The interplay between bending-like ripples and intrinsic body strains can be understood from the underlying energetics at these intermediate widths. Approximating the saddle morphology as $ f(y)\approx \delta_x + [1-\cos (\pi y/w)] \delta_y$ with a transverse amplitude constant much smaller than that along the ribbon, $|\delta_y|\ll \delta_x$, the bending energy due to the two curvatures $\kappa_{xx}\sim -k^2 \delta_x^2 \sin{kx}$ and $\kappa_{yy} \sim \delta_y/w^2 \sin kx$ scales as $\mathcal{E}_b\sim~Dw(k^4\delta_x^2+\delta_y^2/w^4)$. The stretching energy has contributions from both the core and the edge, $\mathcal{E}_s\sim\tau_ek^2\delta_x^2$ and $\mathcal{E}_s^e\sim -\tau_ek^2(\delta_x+\delta_y)^2$. Then, the leading order term for the total energy for stretching is $\mathcal{E}_s+\mathcal{E}_s^e\sim -\tau_ek^2\delta_x\delta_y$. The relation between the two amplitude constants follows from the zero net moment across the ends, which reduces to
\begin{equation}
\label{eq:momentBalance}
-Dw\kappa_{xx} + T \delta_x \sin kx + 2\tau_e (\delta_x + \delta_y) \sin kx \approx 0,
\end{equation}
where the longitudinal force $T=-2\tau_e$. Simplifying, we get the ratio $\delta_y/\delta_x \sim -(kw)^2 / (\tau_ew/D)$ which is also related to the intrinsic (scaled) strain, $\delta_y/\delta_x\sim (kw)^2/(\epsilon_{xx}^0Sw^2/D)$. Substituting in the balance, $\mathcal{E}_b=\mathcal{E}_s + \mathcal{E}_s^e$ yields the observed dependence, $\tau_e^\ast w/D \sim-1$.

%\begin{figure}[tb]
%\includegraphics[width=0.7\columnwidth]{fig4.eps}
%\caption{\label{fig:fig3} (a) Comparison of the analytically predicted stability of of bending-like and twist-like shapes and the buckled shapes observed in atomic-scale simulations for a-GNRs, now plotted as the wave number $k$ versus width $w$. Also shown are representative atomic-configurations for symmetrically rippled ribbons with widths $w=0.72$\,nm, $w=1.93$\,nm and $w=5.12$\,nm.}
%\end{figure}

The stability space for extrinsically strained nanoribbons is shown in~\ref{fig:fig3}a and~\ref{fig:fig3}b for bending-like and twist-like rippling, respectively. Note that the compressive strains $\epsilon_{xx}<0$ result in a different class of solutions as detailed in SI. 
%of the form
%\begin{equation}
%\label{result_negative1}
%f=A_0\cosh{py}+A_1\sinh{qy}+B_0\cos{qy}+B_1\sin{qy}
%\end{equation}
%where the variables $p$ and $q$ are now defined as $p^2=k^2(\xi+1)$ and $q^2=k^2|\xi-1|$.  Again, we recover the relaxed solutions in the limit $\epsilon_{xx}=0$ (see SI). 
The contours in the plots correspond to critical (scaled) edge stresses as a function of (scaled) applied strain and wavenumber. The iso-(edge)stress contour for classical Euler buckling ($\tau_e^\ast=0$) and iso-strain plots for intrinsically strained nanoribbons ($\epsilon_{xx}=\epsilon_{xx}^0$) are also plotted. For small strains and large $kw\gg1$ characterized by non-interacting rippled edges ($\delta_y\sim0$), the critical strain decreases linearly with wavenumber. Here, the bending energy due to the rippled edges is constrained to a penetration width $w_l\sim1/k$~\cite{nr:ShenoyZhang:2008} and therefore scales as $\mathcal{E}_b\sim Dk^4w_l\,\delta_x^2\sim Dk^3\delta_x^2$. Both the ribbon core and the edge contribute to the stretching energy, which scales as $\mathcal{E}_s+\mathcal{E}_s^e\sim -(\epsilon_{xx}Sw+2\tau_e) k^2\delta_x^2$. Equating the two yields the linear dependence, i.e. $\epsilon_{xx}^\ast Sw^2/D \sim - (k^\ast w + \tau_e^\ast w/D)$. At small wavenumbers $kw\ll1$, on the other hand, the critical applied strain is proportional to the edge stress and is independent of the wavenumber for both morphological classes. This follows from the analytical solutions as well as scaling analyses for intrinsically strained ribbons where the interplay between transverse and longitudinal curvatures results in a critical edge stress $\tau_e^\ast w/D$ that is independent of the wavenumber (\ref{fig:fig2_1}); the critical strain varies similarly as evident from the stability diagrams.\footnote{The behavior for $\tau_e\rightarrow0$ should approach that for classical Euler buckling. For $kw\ll1$, the critical strains for bending-like shapes deviate and exhibit a quadratic dependence with $kw$ while those for twist-like rippling are wavenumber independent. To see this, note that the critical axial force for buckling of a simply supported column is $S\epsilon_{xx}^\ast=\pi^2 E I_z/L^2\sim k^2Dw$
%(1-\nu^2) k^2Dw$ 
and $S\epsilon_{xx}^\ast=EI_{xx}A/[2(1+\nu)(I_y+I_z)]\sim D/w$,
%24D(1-\nu)/w$ 
respectively for axial-flexural and axial-torsional buckling. The expressions are determined in the limit $w/h\rightarrow\infty$ with axial moment of inertia $I_z = wh^3/12$,  $I_y = w^3h/12$ and the torsional moment of inertia $I_{xx} = h^3w/3$.
}

The transition behavior at intermediate wavenumbers $kw\sim1$ is characterized by peaks in the critical strains for ribbons subject to tensile and compressive strains. The peaks are apparent in the plot for bending-like ripples and appear at strains larger than the range shown in the plot for twist-like ripples. \ref{fig:fig3}c shows the iso-stress stability curves for the specific case of $w=10$\,nm wide a-GNRs (edge stress $\tau_e=-10.5$\,eV/nm). The wavenumber is plotted on a log-scale to highlight the relative stability of the two morphological classes at small wavenumbers. At applied strains less than the peak strain we see a bifurcation into an additional morphology with smaller wavenumbers. The double-well profiles associated with this sub-class ($k^\ast w=7.5$) are shown in \ref{fig:fig3}d for a-GNRs subject to a uniaxial strain $\epsilon_{xx}=1\%$. The co-stable morphology with larger critical wavenumber exhibits ripples localized to the edges ($k^\ast w=37$), as expected. At much lower strains, we recover the saddle-shapes analyzed earlier. In the case of a-GNRs, the transition occurs at strains below $\approx0.3\%$. As confirmation, the bending-like rippled morphology for $\epsilon_{xx}=0.15\%$ and $k^\ast w=0.5$ is shown in~\ref{fig:fig3}d. 

The additional double-well shapes are also observed in our computations. In fact, transversely rippled center-line morphology is clearly visible in Fig.~\ref{fig:fig1}d for bending-like ripples in ribbons subject to tensile strains, $\epsilon_{xx}=1\%$.  It is interesting to note that these profiles are more susceptible to twist-like rippling as the ribbon midline is largely unmodified compared to bending-like ripples. In fact, for a range of applied strains (0.1\%-0.3\% for the a-GNRs), the double-well shaped rippling can only occur in a twist-like fashion (\ref{fig:fig3}c). Similar peaks also occur in critical strains for ribbons with tensile edge stresses and subject to compressive applied strains. The behavior is consistent with classical Euler buckling of ribbons appropriately modified due to the edge stress; the more general morphology of ribbons subject to tensile edge stresses is discussed in the next section. 

\section{Strained Ribbons with Tensile Edge Stress}
Tensile edge stresses are usually driven by reconstructions and therefore can vary significantly. Recent calculations indicate that this is indeed the case in graphene nanoribbons. a-GNRs with $5-6$ reconstructions result in large stresses, $\tau_e=24$\,eV/nm while those in the $5-7$ reconstructed z-GNRs are almost negligible, $\tau_e=0.02$\,eV/nm. 
%Hydrogenation further modifies the edge stress with $\tau_e=-2.2$\,eV/nm and $\tau_e=15$\,eV/nm for a-GNRs and z-GNRs, respectively~\cite{nr:ReddyShenoyZhang:2009}. 
%As mentioned earlier, these nanoribbons will exhibit a different set of morphological classes as the ribbon core is forced to seek viable out-of-plane displacements. 
The periodic ripples in the suspended nanoribbons can again be categorized broadly into bending-like and twist-like, and both classes have been observed in simulations (\ref{fig:fig1}e). The solutions for the rippled shapes in the small amplitude limit follow directly from \ref{eq:sinusoidalPert}-\ref{eq:rippledBC} and are again detailed in SI (Eq. S7)

\begin{figure}[thb]
\includegraphics[width=0.7\columnwidth]{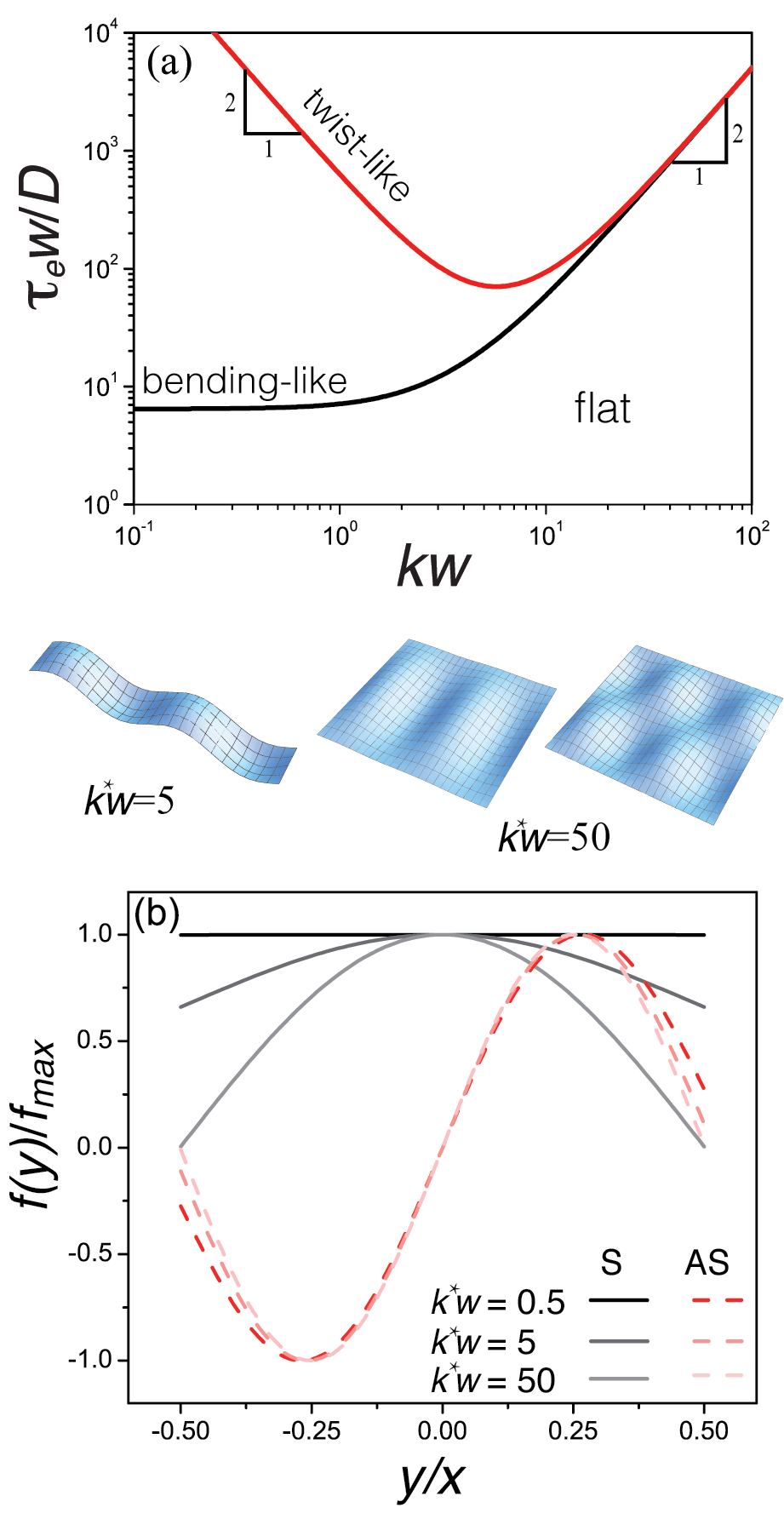}
\centering
\caption{(a) Stability diagram and (b) ribbon profiles as in~\ref{fig:fig2_1}, but for intrinsically strained nanoribbons with tensile edge stresses. For clarity, the dimpled morphologies predicted by our analysis for $k^\ast w=5$ and $k^\ast w=50$ are shown schematically.
\label{fig:tensileEdgeStress}
}
\end{figure}

We restrict our analysis to compressively strained ribbons as the morphology of unstrained or stretched ribbons is trivial - they are always flat since the edge stress is completely absorbed by the relevant end conditions. The reasoning is consistent with our analysis in that the critical contours for rippling shown in Fig.~\ref{fig:fig3}a-b always lie below the $\epsilon_{xx}=0$ line; the rippling requires a compressive stress. For intrinsically strained nanoribbons, the iso-strain contour line for bending-like ripples is plotted in Fig.~\ref{fig:fig3}a (solid white line below $\tau_e>0$) while the detailed stability diagram $\tau_ew/D$ vs. $kw$ and the associated ribbon profiles are plotted in~\ref{fig:tensileEdgeStress}. At large wavenumbers the edges do not interact and past the critical point the core develops dimples around the midline (positive Gaussian curvature) to accommodate the compressive strain. The limit corresponds to Euler buckling of the ribbon core with a scaled critical edge stress that varies quadratically with the critical wavenumber, as expected. The edge interaction at smaller wavenumbers enhances the relative extent of the dimples, i.e. the transverse curvature increases. In the limiting case $kw\ll1$, the critical wavenumber is independent of the edge stress. The trend is qualitatively similar to that in ribbons with compressive edge stress (\ref{fig:fig2_1}a) for the simple reason that in both cases the bending energy is modified by transverse curvatures, albeit opposite in sign. 

The twist-like rippling in these ribbons consists of dimples that alternate across the ribbon midline, corresponding to an axial-torsional buckling mode of the ribbon core. At small wavenumbers (width), the critical edge stress is expectedly larger due to the significant transverse bending (Gaussian) energy associated with the morphology. Increasing the wavenumber reduces this energy and the critical stress decreases quadratically. In the large wavenumber limit, the edges no longer interact and the critical curve is identical to that for bending-like ripples, i.e. the critical stress {\it increases} quadratically with wavenumber. At intermediate wavenumbers, therefore, we see a minimum in the critical edge stress. Note the ribbon profiles are relatively insensitive to the wavenumber. The scaled edge stress for nanometer wide a-GNRs fall within the proximity of the minimum. Although the bending-like ripples are favored, the energetic difference is small in that we see twist-like ripples in some of the computed morphologies (\ref{fig:fig1}e). Finally, the stability with respect to extrinsic compressive strains is shown in~\ref{fig:fig3}a-b. Larger compressive strains always favor bending like ripples since the rippled mid-line is more efficient in absorbing the imposed strain. Of note is a peak critical strain for bending-like ripples that reflects the transition into a higher order Euler buckling mode (in this case, an S-shaped profile) at larger wavenumbers. In fact, there are several such peaks that emerge at larger compressive strains and wavenumbers as the higher buckling modes become viable. 

\section{Concluding Remarks}
The final ribbon morphology has ramifications for electronic properties that are strongly coupled to the ripples and the associated strain distribution, ranging from conductance fluctuations to band gaps. Clearly, the interplay between ribbon structure, geometry and externally applied strains results in a morphological stability space that is considerably richer than the classical edge ripples that are normally associated with these crystalline nanoribbons. For example, tensile edge stresses tend to destabilize the ribbon core while forcing the edges to be flat, and the dimpled morphology affects both equilibrium and transport properties. We note that the excellent agreement in the trends observed in the predicted morphologies and those observed in atomic-scale simulations in graphene nanoribbons is limited to low temperatures as we have ignored the effect of thermal fluctuations. This is to be expected as the phonon fluctuations that couple the in-plane and out-of-plane displacements renormalize the ribbon stiffnesses such they become inherently size dependent. The stability diagrams extracted here are still applicable in that the dominant effect of these fluctuations is to change the scaled edge stresses and strains. In the case of nanoribbons, though, there is an inherent anisotropy due large differences in the thermal fluctuations along the transverse and longitudinal directions. As an interesting example, in systems where the base bending rigidity increases, the twist-like dimples can stabilize at high enough temperatures as the rigidity is enhanced preferentially along the ribbon, $D_x>D_y$. In that case, a statistical understanding of the effect of the resultant shape fluctuations on the ribbon properties becomes necessary.

{\bf Acknowledgements}: This work was supported by the National Science Foundation Nanoscale Science and Engineering Center (NSEC) for High-rate Nanomanufacturing (NSF grant- 0425826) and the Structural Metallics Program, Office of Naval Research, Award No. N00014-06-1-0207.

\section{Methods}

\subsection{Atomistic Simulations}
The computations are performed on graphene nanoribbons with a fixed length, $l=21$\,nm. For extrinsically strained ribbons, the strain was applied to the fixed ends of the ribbons. The width was systematically varied in the range $w=1.2-10$\,nm. The effect of compressive and tensile edge stresses was studied using pristing and reconstructed graphene edges, for armchair and zig-zag edge structures. In each case, the nanoribbons were perturbed by a combination of random and sinusoidal displacements and then relaxed to their equilibrium shape using a conjugate gradient algorithm with an energy tolerance of $10^{-10}$\,eV.

\subsection{Composite elastic framework}
The nanoribbon is approximated as an isotropic elastic thin plate with thickness $h$, width $w$, and length $l$ $({h}\ll{w}\ll{l})$. Our composite framework consists of edge, modeled as a stretched or compressed elastic string that is glued to the ribbon. The edge stress $\tau_e$ is the main material parameter associated with the string, it has no bending stiffness. Additionally, the edge is assumed to be sharp such that that the width of the ribbon is in fact that of the ribbon core. 

\subsubsection{Governing Equations}
The governing equations follow from standard Fv-K equations in the small deflection limit,  
%\begin{equation}
%\label{von1}
%\nabla^4{\Phi}=-\frac{1}{2}S\diamond^4(\zeta,\zeta), \qquad D\nabla^4{\zeta}=\diamond^4(\Phi,\zeta).
%\end{equation}
\begin{equation}
D\nabla^4{\zeta}=\Phi_{,yy}\zeta_{,xx} \qquad
\nabla^4{\Phi}=-S(\zeta_{,xx}\zeta_{,yy} - \zeta_{,xy}^2).
\end{equation}
Here $\Phi$ is the Airy function associated with the in-plane stresses and the operator $\nabla^4A=A_{,xxxx}+2A_{,xxyy}+A_{,yyyy}$. 
For long ribbons, end effects can be safely ignored; the ribbon has negligible transverse and shear stresses, and the governing equation simplifies to
\[
D({\zeta}_{,xxxx}+2{\zeta}_{,xxyy}+{\zeta}_{,yyyy})=T{\zeta}_{,xx}
\]
where $T=S\epsilon_{xx}$ is the net longitudinal force. Setting our origin on the midline, the force balance modified by the  edge stresses and free torques specify the boundary conditions,
\begin{eqnarray}
\{\zeta_{,yy}+\nu\zeta_{,xx}\}|_{\pm{\frac{w}{2}}}&=&0,\nonumber\\
\left \{\zeta_{,yyy}+(2-\nu)\zeta_{,yxx} \pm \frac{\tau_e}{D}\zeta_{,xx} \right \}|_{\pm{\frac{w}{2}}}&=&0.\nonumber
\end{eqnarray}
Substituting the sinusoidal deflection $\zeta(x,y)=f(y)\sin{kx}$ in the equations yields the eigenvalue problem for the ribbon shape in terms of the ribbon strain $\epsilon_{xx}$, Eqs.~2 and Eq.~3.

\subsubsection{Analytical Solutions}
The parameter sets considered here are yield fully analytical solutions for the ribbon morphology. In particular, the form of the solutions differs for relaxed and strained ribbons, as detailed in Supplementary Information (SI).

%the out-of-plane coupling is driven by residual stresses along the longitudinal direction. In this small deflection limit, the Fv-K equations reduce to (see SI) 
%\begin{equation}
%\label{von1}
%\nabla^4{\Phi}=-\frac{1}{2}S\diamond^4(\zeta,\zeta), \qquad D\nabla^4{\zeta}=\diamond^4(\Phi,\zeta).
%\end{equation}
%\begin{equation}
%\label{eq:von1}
%D\nabla^4{\zeta}=\Phi_{yy}\zeta_{xx} \qquad
%\nabla^4{\Phi}=-S(\zeta_{xx}\zeta_{yy} - \zeta_{xy}^2)
%\end{equation}
%where $\Phi$ is the Airy function associated with the in-plane stresses and the operator $\nabla^4A=A_{xxxx}+2A_{xxyy}+A_{yyyy}$. 
%Setting our origin on the ribbon centerline, the edge forces together with free torques specify the boundary conditions,
%\begin{eqnarray}
%\label{eq:BCs}
%\{\zeta_{yy}+\nu\zeta_{xx}\}|_{\pm{\frac{w}{2}}}&=&0,\nonumber\\
%\{\zeta_{yyy}+(2-\nu)\zeta_{yxx}] \pm \frac{\tau_e}{D}\zeta_{xx}\}|_{\pm{\frac{w}{2}}}&=&0.
%\end{eqnarray}

\widetext{

\pagebreak

\begin{center}
{\bf Supplementary Information for \\
``Rippling Instabilities in Suspended Nanoribbons"}

{\bf Hailong Wang and Moneesh Upmanyu\\
%\email{mupmanyu@neu.edu}
\rm Group for Simulation and Theory of Atomic-Scale Material Phenomena (stAMP), \\
\rm Department of Mechanical and Industry Engineering, and Bioengineering, \\
\rm Northeastern University, Boston, Massachusetts 02115, USA.
}

\end{center}
}

For each parameter set $(\tau_e, \epsilon_{xx})$, our composite framework yields fully analytical expressions for the critical ribbon shapes. The solutions for the parameter sets analyzed in the text are detailed below. 

\section{Zero strain $\epsilon_{xx}=0$}
The governing equation for relaxed suspended nanoribbons reduces to 
\begin{equation}
\tag{S1}
f_{,yyyy}-2k^2f_{,yy}+ k^4 f =0.
\end{equation}
The general solution for both bending-like and twist-like ripples is of the general form
\begin{equation}
\tag{S2}
f=\sum_{n=0,1} A_n(ky)^n \cosh{ky} + B_n (ky)^n \sinh{ky}, 
\end{equation}
where the set of constants $(A_0, B_1)$ and $(A_1, B_0)$ correspond to bending-like (symmetric, S) and twist-like (anti-symmetric, AS) ripples, respectively. The constants can be evaluated using the boundary conditions (Eq.~5). The solution also yields the critical buckling point and the associated shape function. For bending-like ripples, we get
\begin{equation}
\tag{S3}
\frac{{\tau_e^\ast}w}{D}=-\frac{(1-\nu)k^\ast w[(3+\nu)\sinh{k^\ast w}-(1-\nu)k^\ast w]}{2(\cosh{k^\ast w}+1)}
\end{equation}
At small wavenumbers, $kw\ll1$, ${\tau_e^\ast}w/D=-(1-\nu)(1+\nu)(k^\ast w)^2$ and for $kw\gg1$, ${\tau_e^\ast}w/D=-(1-\nu)(3+\nu)k^\ast w/2$.
The associated shape function is, 
\begin{equation}
\tag{S4}
f=A\{\frac{(1-\nu)kw\tanh{(kw/2)}+4}{2(1-\nu)}\cosh{ky}-ky\sinh{ky}\}
\end{equation}
where $A$ is an arbitrary constant.
The corresponding relations for twist-like ripples are
\begin{equation}
\tag{S5}
\frac{{\tau_e^\ast}w}{D}=-\frac{(1-\nu)k^\ast w[(3+\nu)\sinh{k^\ast w}+(1-\nu)k^\ast w]}{2(\cosh{k^\ast w}-1)}
\end{equation}
Again, for $kw\ll1$, ${\tau_e^\ast}w/D=-4(1-\nu)$ and for $kw\gg1$, ${\tau_e^\ast}w/D=-(1-\nu)(3+\nu)k^\ast w/2$.
The solution for the ribbon profile is
\begin{equation}
\tag{S6}
f=A\{\frac{(1-\nu)kw\coth{(kw/2)}+4}{2(1-\nu)}\sinh{ky}-ky\cosh{ky}\}
\end{equation}

\section{Tensile strain $\epsilon_{xx}>0$}
The general solution for tensile strains $\epsilon_{xx}>0$ is
\begin{equation}
\tag{S7}
f=A_0\cosh{py}\cos{qy}+B_0\cosh{py}\sin{qy}+A_1\sinh{py}\sin{qy}+B_1\sinh{py}\cos{qy}.
\end{equation}
Here $p=k\sqrt{(\sqrt{1+k_r^2/k^2}+1)/2}$ and $q=k\sqrt{(\sqrt{1+k_r^2/k^2}-1)/2}$, where $k_r$ is defined as $k_r^2=T/D$. As a check, the solution for $\epsilon_{xx}=0$, Eq.~6, can be obtained in the limit $k_r{\ll}k$ and by letting $\cosh{py}{\rightarrow}\cosh{ky}$, $\sinh{py}{\rightarrow}\sinh{ky}$, $\cos{qy}{\rightarrow}1$ and $\sin{qz}{\rightarrow}ky$.

\section{Compressive Strain, $\epsilon_{xx}<0$}
In the case of compressive strains, the general solution for $k_r<k$ is 
\begin{equation}
\tag{S8}
f=A_0\cosh{py}+B_1\sinh{qy}+A_1\cosh{qy}+B_1\sinh{qy}
\end{equation}
Here $k_r^2=-T/{D}$, $p=k\sqrt{1+k_r/k}$, $q=k\sqrt{1-k_r/k}$. For small wavenumbers, the modified solution takes the form
\begin{equation}
\tag{S9}
f=A_0\cosh{py}+B_0\sinh{qy}+A_1\cos{qy}+B_1\sin{qy}
\end{equation}
where $p=k\sqrt{k_r/k+1}$, $q=k\sqrt{k_r/k-1}$. For special case $k=k_r$,
\begin{equation}
f=A_0\cosh{\sqrt{2}ky}+B_0\sinh{\sqrt{2}ky}+A_1+B_1y.\nonumber
\end{equation}

\section{Intrinsically Strained Ribbons, $\epsilon_{xx}=-2\tau_e/S$}
The general eigenvalue solution for generally strained ribbons can be written as a function of scaled wave number $kw$, scaled edge stress ${\tau_e}w/D$ and scaled strain $Tw^2/D$,
\begin{equation}
\tag{S10}
\frac{\tau_e^\ast w}{D}=\Psi(k^\ast w, T^\ast w^2/D)
\end{equation}
For freestanding nanoribbons, we can obtain the solution by substituting $\epsilon_{xx}=-2\tau_e/Sw$ into Eq.~S10. The results, $\tau_e^\ast w/D=\Phi(k^\ast w)$, are shown graphically as white line in Figs. 3a and 3b of main text. Due to the complicated mathematical formulae for $\Psi(k^\ast w, T^\ast w^2/D)$ and $\Phi(k^\ast w)$, $\tau_e^\ast w/D$ can be only evaluated numerically.

%\begin{thebibliography}{99}
%\end{thebibliography}

\end{document}